\documentstyle[prd,aps,floats]{revtex}

\def\be{\begin{equation}}
\def\ee{\end{equation}}
\def\ben{\begin{eqnarray}}
\def\een{\end{eqnarray}}

\begin{document}

\input epsf
\renewcommand{\topfraction}{1.0}
\twocolumn[\hsize\textwidth\columnwidth\hsize\csname 
@twocolumnfalse\endcsname

\title{Phenomenological model of the weak interaction}

\author{Franz E.~Schunck\footnote{E-mail: fs@thp.uni-koeln.de}}

\address{Institut f\"ur Theoretische Physik, Universit\"at zu
K\"oln, 50923 K\"oln, Germany}

\date{\today}

\maketitle

\begin{abstract}
We use the informations known so far about elementary particles
in order to construct a simple model. We find a reason for the gyromagnetic
factor of 2 for leptons and a vivid imagination for the weak interaction.
By this, we understand, why the elementary particles with lowest mass
are stable and all other unstable.
\end{abstract}





\vskip2pc]

\section{Introduction}

The attempt to understand the internal structure of the electron has long
tradition and goes back to Hicks pure geometrical idea of vortices \cite{hicks}
but without explicit relation to the electron.
A few years later, already in 1907, the electron as ringlike axial structure
was introduced, i.e., electromagnetic energy circles around an axis
\cite{stark}. In \cite{lorentz}, the idea of a deforming electron is given
the magnetic moment of which stays nevertheless constant.
Using the experimental results of Stern and Gerlach \cite{stern}, Uhlenbeck and
Goudsmith \cite{uhlen} introduced the idea of an intrinsic angular momentum
of the electron which is twice as large as one would expect classically
and results in the Land\`e or g-factor. Dirac \cite{dirac} followed from the
linearised relativistic Schr\"odinger equation for an electron in a magnetic
field that both an anti-particle (the positron) should exist and
the exact value of the magnetic moment of the electron.

There are a lot of theories which try to understand the internal strcuture
of elementary particles in general. There exist the ideas of screwvortices
(the {\em archon}) \cite{wiener}, left and right archons together
form a particle.
Broglie applied this as {\em m\'ethode de fusion} \cite{broglie}.
However, Born and Peng introduced a basic element called
the {\em apeiron} \cite{born}.
H\"onl \cite{hoenl} introduced an internal motion of the
electron and Schr\"odinger \cite{schroedinger} a trembling motion.
H\"onl stressed that an electron could be a {\em pole-dipole} particle.
Jehle considered elementary loops which in superposition form electron or quark
\cite{jehle}. Pekeris used a hydromechanical model with stationary circulations
\cite{pekeris}. Dahl had the idea of a rotor model with two elements
\cite{dahl} whereas Hughston applied two twistor elements \cite{hughston}.

Following Mack and Petkova \cite{mack}, quarks can be described by
condensed vortices. 
Bopp \cite{bopp} followed from the interaction of photons with matter
that photons could not consist of electron pairs which divide up due to
the impact. Harari \cite{harari} introduced as basic element the {\em rishon}
so that elementary particles consist of three of them. He concluded that
an electron cannot consist of several rishons because the total energy
of the electron has to be smaller than the one of its constituents.
Pavsic et al.~\cite{pavsic} uses Cliffords $4\times 4$ matrix and interprete
their result as pointlike charge the orbit of which is a cylindrical helical line.
Following Hautot \cite{hautot}, charge and magnetic moment are two objects
of the eigenstructure, spherically for matter and charge,
axially for angular and magnetic moment.
Wasserman \cite{wasserman} presents as electron model a M\"obius band
asymmetrically cut so that two stripes become interlinked, a M\"obius band
becomes interlinked with a triple twisted M\"obius band. In his opinion,
the twisting is correlated with the spin, and the interlinking with mass
and charge.

\section{Particle characteristics}

In this work we are considering some of the characteristics of elementary
particles. Neutrinos take a special role due to their electrical neutrality.
All other particles have an
electric charge of $\pm$1, $\pm$2/3, or $\pm$1/3 of the elementary charge $e$
if we take into account
the anti-particles as well. The electric charge is conserved.
Every particle has a mass (i.e.~its gravitational charge) which is conserved
as well. The experiments have revealed a weak charge where no conservation law 
exists. In this work, we ignore the influence of the strong interaction;
leptons do not take part in the latter one.

Furthermore, characteristics like the magnetic dipole are determined,
but its exact values are still unclear but for the electron and the muon
\cite{pdg}. It has been taken quite some time and a lot of
experimental support to develop a model for the electromagnetic and weak
interactions, which is now known as the standard model of the electroweak
interaction. From this, we understand that all fermions take part in the
electromagnetic interaction independent on its spin, but only fermions with
left handed spin interact weakly; this is known as parity breaking of
the weak interaction. A reason for the latter has not been found so far.

The following particles (and its anti-particles) decay by the weak
interaction the more rapidly,
the heavier they are: the leptons $\mu$, $\tau$, and the quarks
$d$, $s$, $c$, $b$, $t$. For example, one of the $d$ quarks inside a free neutron
needs about 900 seconds in order to decay, whereas the tauon $\tau$
decays almost immediately within $10^{-13}$ seconds.
In a weak decay, the interaction can be described by four fermion fields which
take part in it (besides an intermediate boson); there is no interaction between
two particles, but there is one particle ``waiting'' for the interaction with
its three potential decay products.

There is another problem within the standard model of the electroweak
interaction: All fermions have mass zero. One needs the so-called Higgs
mechanism in order to introduce masses for the elementary particles
\cite{ryder,itzykson}.
A Higgs particle which is so far not found experimentally gives every
elementary particle its mass by a symmetry breaking process.

An electron has an intrinsic magnetic moment which so far can only be understood
by a quantum mechanical treatment. The magnetic moment of the electron is
\be
{\bf \mu_s} = -g_s \mu_B {\bf s}/\hbar  \label{mu}
\ee
where $\mu_B$ is Bohr's magneton and $g_s$ the Land{\'e} factor which
is equal to 2 in Dirac mechanics (but slightly larger due to quantum
electrodynamical effects).
It is related to the intrinsic angular momentum of the particle.
We notice that $\bf \mu$ is antiparallel to the spin angular momentum.
Classically, we can imagine that the spin angular momentum is produced
by the electron mass spinning around some axis whereas the electric current
produces a revolution in the opposite direction due to
the negative charge of the electron.

In the next section we introduce the basics of our model before we start
to go into the details for leptons, quarks, and neutrinos.
Then, in Section \ref{mass} we discuss critically our model and finish
the article by some remarks for future work.

\section{The model} \label{model}

We have to distinguish between a model for the leptons and two for the quarks.
Separately, we discuss the situation for neutrinos afterwards.
We outline here only shortly our assumptions.

\noindent
{\bf Electric charge}

$\bullet$ Leptons with negative electric charge:\\
A charge of the amount of two negative elementary charges surrounds
a charge of the amount of one positive elementary charge. Both move around
a common center of mass \cite{schunck}. (We do not assume anything about the structure
of these particles, i.e., whether they are bosons or fermions.)

$\bullet$ Quarks with electric charge $+2/3$:\\
A charge of the amount of 4/3 of a positive elementary charge surrounds
a charge of the amount of 2/3 of a negative elementary charge. Both move around
a common center of mass.

$\bullet$ Quarks with electric charge $-1/3$:\\
A charge of the amount of 4/3 of a negative elementary charge surrounds
a charge of the amount of 3/3 of a positive elementary charge. Both move around
a common center of mass.

(Analogue for the anti-particles.)

\noindent
{\bf Mass}

$\bullet$ Leptons and Quarks\\
We assume that the mass of the Higgs particle is distributed inside the
elementary particle. We take: $m_{Higgs}=2.5 \times 10^5 m_e = 127.75$ GeV
where $m_e$ is the mass of the electron. (Notice:
This is not the mass of the elementary particle as measured from outside.)

\noindent
{\bf Weak interaction}

$\bullet$ Leptons and Quarks\\
We assume that a dielectric medium simulates the weak interaction.
The dielectric is as usually given by a dielectric constant $\epsilon_r$,
hence, we assume a homogeneous medium.

\section {Leptons} \label{leptons}

Here, we analyse the consequences of the model given in the last section.
We start with the leptons.

The sum of the two charges inside the particle gives one negative elementary
charge as found experimentally; we repeat here shortly the results of the model
given in \cite{schunck}. Two opposite charges means that our model
possesses a magnetic dipole moment as it is measured for $e, \mu, \tau $.

Due to the internal structure of our model, we can derive an intrinsic
magnetic dipole moment \cite{schunck}.
We assume that the double negative charge $-2e$ moves around the single
positive charge $+1e$, so that observed
from outside we will measure the movement of this double charge.
Seen from distance the object possesses the mass $m_x$ where $x$ stands for
electron, muon, or tauon, respectively (and not the Higgs mass). 
Then, the leptonic magnetic dipole moment $\mu$ is exactly Bohr's
magneton, i.e., $\mu = \mu_B = e \hbar / 2 m_x$, cf.~Eq.~(\ref{mu}).
We recognize that the introduction of an
arbitrary factor $g_s$ is no longer necessary; instead, our model
gives the g-factor a physical meaning. It is $g_s=Z_1$, i.e.,
$g_s$ is the amount of charges $Z_1$ surrounding the central charge $Z_2$.
In case of leptons, this is $g_s=Z_1=2$.
Of course, in our simple model, we cannot clarify the exact value of the
gyromagnetic factor given by quantum electrodynamics.

Clearly, the spin of the elementary particle can be described by our model
simply by the imagination that the double charge $-2e$ surrounds
the single central charge $+1e$. Then, two ways of circulation 
perpendicular to the direction of motion are possible,
which suggest spin-up or spin-down, respectively.

Furthermore, we assume that the mass of the Higgs particle is distributed
in some way between the double and the single charge. It is, for the time being, not 
important how this distribution is implemented. What counts is that the
reduced mass of this system is equal to the Higgs mass. Of course, due to the
measured magnetic dipole moment we can derive that the mass is mainly
concentrated in the single charge. The Standard model shows that
without the Higgs mechanism all elementary particles are massless.
Only the assumption that there exists a spontaneous symmetry breaking
resulting in a Higgs particle gives elementary particles a mass.
In a later section we will go in more detail into the issue of mass distribution
inside a particle.

Our model shows that there is a bound system of two opposite charges.
So, up to some point, we can use the results of
the hydrogen atom as known from quantum mechanics.
But, furthermore, we have to take into account the results 
of the experiments.
We are aware of no excited states of elementary particles.
This means that we have to introduce an energy shift in the hydrogen model
so that excited states exist only for positive energy levels.
The first excited state in the hydrogen model with energy $E_2$
has the energy $E_2=E_1/4$ where $E_1$ is the energy of the ground state.
The energy shift means that the ground state energy is changed by a factor $3/4$.

From quantum field theory and from experiments, we know that inside and around
an electron permanently virtual photons and virtual electron-positron pairs
are created; this is known as vacuum polarisation. Due to this, we assume
additionally that our two charges move {\em not} in vacuum but in a
{\em dielectric} which is strongly restricted in space. Hence, bound states
exist only for a small area, beyond this the two charges cannot ``live''.
Because of this finiteness of the dielectric medium, we can cut
the excited energy levels as well; otherwise they would exist also outside the medium.
The jump between two energy levels would only be possible if the two charges
can allow to transmit a spin-1-particle, the photon (spin conservation).
If not, there could also exist more energy levels but these would not be reachable.

In a nutshell, we sum up our results. We have two charges inside a finite
dielectric medium; they build a bound system which can be described
by quantum mechanical means. We use here a spherically symmetric system;
taking a cylindrically symmetric description makes only a slight difference
\cite{fluegge}.
This system can exist in different ways depending
on the characteristics of the dielectric. So, we will use it to form
elementary particles which distinguish in the dielectric constant.
We take as given that there exist three charged leptons.

The energy of the shifted ground state of the two charges
is given by
\ben
E_1 & = & - \frac{1}{2} m_e c^2 \, \frac{\alpha^2}{\epsilon_r^2} \, 
          \frac{m_{Higgs}}{m_e} \, Z^2 \, \frac{3}{4} \\
 & = & - 13.605 \frac{1}{\epsilon_r^2} \, \frac{m_{Higgs}}{m_e} 
       \, Z^2 \, \frac{3}{4} \; {\rm [eV]} \, , \label{E1}
\een
where $m_e$ is the mass of the electron, $\alpha = e^2/(4\pi \epsilon_0 \hbar c)$
the fine-structure constant, $\epsilon_r$ the dielectric constant of the 
dielectric, $m_{Higgs}$ the mass of the Higgs particle, $Z=Z_1 \times Z_2$
the product of the two charges involved, and the shift factor $3/4$.
For leptons we have $Z_1=2$ and $Z_2=1$.
Then, for the electron, we set $E_1=-m_e c^2=-0.511$ MeV so that we can determine the
dielectric constant $\epsilon_r$ of the dielectric; below we go into the details
why we can set this.
We find, $\epsilon_r (e)= 4.47$. A value greater than 1 indicates of a
medium which reduces the electric fields of the two charges. We conclude that
we have a stable energy state.

For a muon, we set $E_1 = 106$ MeV from which we derive
$\epsilon_r (\mu)= 0.31$. In this medium, the electric field of the two
charges are enhanced very strongly. This dielectric
is highly unstable.

For a tauon, we set $E_1 = 1780$ MeV from which we derive
$\epsilon_r (\tau)= 0.0757$.
In this medium, the electric fields are enhanced even stronger than for a muon.
This dielectric is highly unstable.

The general formula for the leptons is
\be
E_1({\rm leptons}) = - 10.20375 \frac{1}{\epsilon_r^2} \; {\rm [MeV]} \, .
\label{emt}
\ee

These results are in best agreement with observations where one finds that
the electron is stable but the muon decays within about $10^{-6}$ s and the 
tauon much quicker within about $10^{-13}$ s. The last can be understood
in our model: The dielectric constant of the muon is closer to 1 than the
one of the tauon, so that the tauon is more unstable than the muon.

\section {Quarks} \label{quarks}

The situation for quarks is not so clear because no free quarks have been
observed so far. We do not have a clear
measurement of a magnetic dipole moment, hence, we cannot derive how many
elementary charges circulate around a central charge. We can suppose that
there exists a magnetic dipole moment of a quark as can be obtained
from theories \cite{koepp}. Because quarks only appear in combinations of 
two or three of them the measurement of the composite can give us hints on
their intrinsic nature. For example, the proton consisting of two
$u$- and one $d$-quark possesses a magnetic dipole moment. However, the
magnetic dipole moment of the neutron (two $d$-, one $u$-quark) is very small,
perhaps it vanishes \cite{oshimo}. A non-vanishing moment for the neutron
is important because it would prove parity symmetry breaking.

Another problem is the mass of the quarks \cite{pdg}, it depends on a model
chosen. We apply here the valence quark model and use the results of
the Particle Data Group. For the u-quark, for the time being,
a lower limit of 1.5 MeV and an upper limit of 3.0 MeV is provided.
For the $d$-quark, it reveals a mass range between 3.0 and 7.0 MeV.
For these light quarks, we take both these values because they are
important in determing the stability of them. For the heavier quarks,
we only use the best known value without checking the range; they are
unstable anyway.

Again, we can use the results of the discussion of our model
in section \ref{leptons}. First, the 2/3-quarks are investigated.
For the $u$-quark, we have $E_1=1.5$ MeV, and so $\epsilon_r (u)= 1.16$;
for $E_1=3.0$ MeV, we find $\epsilon_r (u)= 0.82$. But this means that
within our model and choice of the amount of the two charges,
there is a mass range where the $u$-quark is unstable (but cf.~the 
discussion for the $d$-quark below).
The stability is guaranteed up to a $u$-mass of about 2.016 MeV, i.e.
\be
E_1({\rm 2/3-quarks}) = - 2.016 \frac{1}{\epsilon_r^2} \; {\rm [MeV]} \, .
\label{23}
\ee

Two remarks are in order.
Of course, we could change our choice of the charges. For the product
$Z=2.44$, we
would find $\epsilon_r (u)= 1.0$ for $E_1=3.0$ MeV, so that the
$u$-quark is stable. But this would mean that we had to fine-tune our model
for $Z_1$ and $Z_2$ which would make the model less attractive.
Because quarks have a charge of one or two third of an elementary charge,
it is more likely that the two charges $Z_1, Z_2$ have multiples
of one third of an elementary charge as well.
Secondly, even the upper value of 3.0 MeV is experimentally very imprecise.
Only a few years ago, 4.0 MeV was the upper limit. Furthermore, there
exists theories stating that the u-quark may be even massless \cite{pdg}.

For the $c$-quark, we set $E_1=1250$ MeV, so that $\epsilon_r (c)= 0.0402$.
For the $t$-quark, we set $E_1=170.9$ GeV, so that $\epsilon_r (t)= 0.0034$.
Both particles are highly unstable, as confirmed experimentally and given by
our model. The $t$-quark with the $\epsilon_r$-value closer to 0 has a smaller
lifetime than the $c$-quark.

Now, we continue with the 1/3-quarks. We find the general formula
\be
E_1({\rm 1/3-quarks}) = - 4.535 \frac{1}{\epsilon_r^2} \; {\rm [MeV]} \, .
\label{13}
\ee
This means for the limits of the $d$-quark:
For $E_1=3.0$ MeV, we have $\epsilon_r (d)= 1.23$, and, for $E_1=7.0$ MeV,
we have $\epsilon_r (d)= 0.805$.
For the $s$-quark, we set $E_1=95$ MeV, so that $\epsilon_r (s)= 0.218$.
For the $b$-quark, we set $E_1=4.2$ GeV, so that $\epsilon_r (b)= 0.0328$.
Again, the $s$- and $b$-quark are clearly and highly unstable in our model.

The situation with the $d$-quark is more complicated. The discussion for the
charge product $Z$ given above for the $u$-quark is valid here as well.
Secondly, the experimentally determined mass limits have been decreased also
for the $d$-quark as for the $u$-quark (from 4.0 to 3.0 MeV or 8.0 to 7.0 MeV, 
respectively). In this case, it means that within our model the $d$-quark
has been stabilized at the lower limit. Actually, this is not a problem.
We know that the $d$-quark is stable depending on the circumstances;
a free neutron is unstable, whereas a bound neutron can be stable.
In general, our model might give a hint that quarks may have no fixed mass,
it depends on the physical situation. The mass of an elementary particle
is the result of the dielectric (which is here the weak interaction) which 
is not restricted by a conservation law. For example, the mass of the 
electron inside a solid state and with a strong magnetic field applied
is influenced dramatically \cite{stormer}.

Of course, one could have chosen also other values for the two charges,
e.g., $Z_1=5/3$ and $Z_2=1$ for the 2/3-quarks. But we believe the
values given in section \ref{model} produces the optimal energy values in
Eq.~(\ref{23}) and (\ref{13}).

\section{Neutrinos} \label{neutrinos}

The existence of a magnetic moment of neutrinos is discussed in literature
and upper limits are given \cite{balantekin}. Because we understand
magnetic moment as movement of a charge in our model, we can apply it
to them as well. We have $Z_1=Z_2=1$ and so
\be
E_1({\rm neutrinos}) = - 2.5509 \frac{1}{\epsilon_r^2} \; {\rm [MeV]} \, .
\label{neut}
\ee
Neutrino masses are not very well known. Only since a few years at all, it is even
clear that they possess a mass due to the effect of mass oscillation
which was clarified by the Super-Kamiokande and the SNO experiment
\cite{sno}.
The electron neutrino mass has an upper limit of about 2.2 eV,
the muon neutrino of about 170 keV, and the tau neutrino of
about 15.5 MeV \cite{pdg}. This would mean that the tau neutrino
would be unstable in our model.

\section{Mass distribution} \label{mass}

The model presented in sections \ref{model}-\ref{neutrinos}
is attractive because of the few input parameters which are taken from
what we know of experiments and theories. The results show what is found
in experiments: a particle is stable or unstable with respect to the weak
interaction. 

But if we look into more detail, we have to recognize that the distribution
of mass inside a particle and the total mass of the particle is a 
difficulty which has to be discussed. Because there are two charges in our
bound system, we applied the results of the quantum mechanical model
of the H-atom. Then, we used the formula for the energy level $E_1$ and
assumed that this is equal to the mass of the particle.

In case of the H-atom, the energy value $E_n$ describes the amount of
energy which is removed from the system by a photon.
Furthermore, the reduced mass in the formula is the result of the
mass of the electron
and the proton, and is close to the mass of the electron. From outside,
we measure as H-atom-mass the sum of the masses of proton and electron
and reduce this by $E_n$ if the electron is in the $n$-th energy state.
(We neglect here other quantum numbers.)

In our model, we assumed that the reduced mass is the Higgs mass.
That would mean that the circulating charge has a bit more than the Higgs mass
and that the central charge would weigh, e.g., as in case of the H-atom,
about $10^3$ times that of the circulating mass. Hence, in total, we 
have an object with about $10^3$ of the Higgs mass. This should be
measured as particle mass from outside and is in clear contradiction
with experiment. This is the description of the H-atom where
vacuum dominates between the two charges.

Instead, one could assume that there is only the elementary particle mass
(and no Higgs mass) present which 
is mainly concentrated in the central charge and that the mass of the circulating
charge $m_{Z_1}$ is, for example, one thousandth of the central charge.
For the electron model, we replace $m_{Higgs}$ by $m_{Z_1}=m_e/1000$ and find
a dielectric constant of $\epsilon_r=2 \times 10^{-4}$. Correspondingly,
all other elementary particles have even smaller $\epsilon_r$-values.
We recognize that by correcting the mass distribution, we find
that all particles are unstable due to a dielectric constant smaller than 1
which is, again, in contradiction with observation.

The model presented in section \ref{model} shows that there appears an
additional factor $m_{Higgs}/m_{Z_1}$ in Eq.~(\ref{E1}). We conclude that
the symmetry breaking Higgs mechanism is valid only inside the dielectric
medium and affects in this way that the large amount of the Higgs mass
is {\em not} measured from outside. Beyond the carrier of the dielectric
medium we measure just the electron mass (or mass of the elementary particle
in general, respectively).
Hence, it is just the right value of the mass of the Higgs particle
which is responsible for a stable electron and a stable $u$-quark.

In principle, we use here the elementary particle as the {\em photon}
in the H-atom model. For some unknown reason, we do not measure the
mass which is inside the particle, but just the energy=mass of
our ``photon''. This problem is already well known \cite{harari}.
Following Heisenberg's uncertainty relation, the amount of a
composite system is closely related to the kinetic energy of its parts.
The smaller the system, the larger the kinetic energy of its constituents.
Inside an electron the lower limit for constituents is just about
100 GeV, which is (more or less) exactly the Higgs mass which we used in our
model. Again, as we discussed above already, we have the paradox situation
that the energy of the constituents is much larger than the mass of the
composite system. Of course, this could mean that an electron is
{\em elementary}. But our model can explain its magnetic dipole moment
and the stability of elementary particles with respect to the weak interaction.
We cannot find a solution here.
Clearly, we need more discussions on this point in future.

\section{Discussion}

In this article, we have tried to calculate the mass of elementary particles.
We introduced a new structure: two opposite charges; we do not make any
assumptions on the nature of these charge elements. For leptons, this
leads to the explanation of the magnetic dipole moment, and hence, the spin.
We calculated the bound system quantum mechanically. We assumed that
the two charges ``live'' in a dielectric medium due to vacuum polarization
effects. In theory, elementary particles are massless, just
a symmetry breaking mechanism leads to a mass where a heavy Higgs particle
is introduced which we used in our model.
Then, we have everything to calculate the elementary particle mass.
We do this by determining certain values of the dielectric constant.
We confirm the stability of the elementary particles with respect to the weak
decay. We conclude that the weak interaction can be interpreted
as dielectric medium.

There has been great effort to investigate electrons in two-dimensional
systems at low temperature exposed to a high magnetic field
\cite{stormer,yoshioka}.
The electrons are inside a solid and start to behave completely
different. It seems that they have changed their internal structure,
they have a charge smaller than the original one. The attached magnetic field
changes the characteristic of the electron from fermionic to bosonic
and back to fermionic depending on the field strength. The mass of the
new object is unrelated to the original mass of the electron. From
experiments, one is sure that the electron has not split up into pieces.
Of course, all of this can be well understood in the context of the
fractional quantum Hall effect. A fractional
charge as in the two-dimensional electron system can only be explained
if we assume that our two charges can change their charge.

In our model, we give a new picture how the interior of an electron
looks like. It can be expected that there is an interaction between the
two charges and the dielectric. For example, in case of a macroscopic 
dielectric by applying a magnetic field the dielectric constant
can be changed \cite{voss}.
In our model, this means that the mass of the electron changes.

\section*{Acknowledgments}

We would like to thank Eckehard W.~Mielke for discussions.

\end{document}